# Comprehensive Three-dimensional Computational Model Enables Design of Nanostructured Infrared Detectors

*Dingkun Ren*

Department of Electrical and Computer Engineering, University of California, Los Angeles, Los Angeles, California 90095, United States

[*]Email: dingkun.ren@ucla.edu

# ABSTRACT

Due to the unique three-dimensional (3-D) geometries of nanowires—i.e., large surface-to-volume ratios and smaller cross-sections at the nanowire-substrate interfaces—their carrier dynamics are much more complicated than those of thin films. Therefore, analytical solutions cannot be found for these nanostructures and a more comprehensive scheme of 3-D modeling is necessary to interpret their intrinsic carrier dynamics. To date, most modeling studies for nanowires have focused on electromagnetic properties (e.g. optical modes and optical absorption). However, very few studies have combined optical and electrical simulations together to probe the temporal and spatial carrier motions within nanowires. In this work, we present a comprehensive nanowire optoelectronic transient model and photoresponse model, allowing us to investigate carrier lifetimes and their fundamental correlations with material properties, as well as responsivities and detectivities for nanowire-based optical devices for photodetection (i.e., photodetectors). We believe this work can stimulate further experimental and theoretical work and unveil the real strength of 3-D computational models for exploring carrier dynamics in nanowires and nanostructured materials.

Technological advances in optical sensing have led to a rapid development of high-performance semiconductor photodetectors and their focal plane arrays [1-3]. Particularly, there is an increasing interest in developing "miniature" pixels in a few micron or submicron scale for high-resolution, highly compact, and low-noise imaging systems. To optimize optical/electrical performance for those "small" devices, it is important to understand carrier dynamics in those structures and properly simulate photocurrent, or photoresponse, to estimate their optical performance. However, it is challenging to do so. This is because such structures have much larger surface-to-volume ratios and smaller cross-sections at the mesa-substrate interface, and thus analytical solutions cannot be found [4-6].

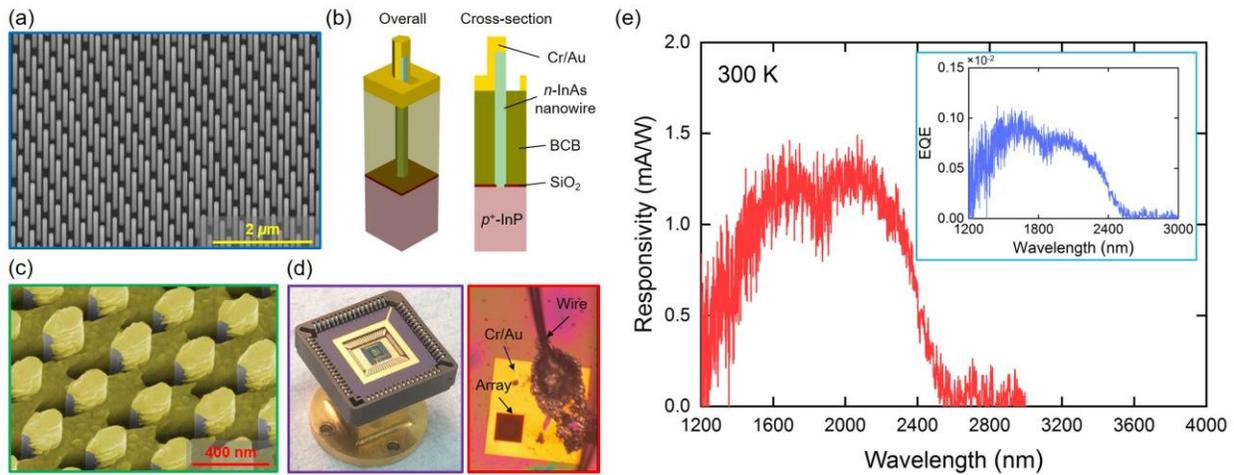

**Figure 1.** SWIR InAs nanowire photodetectors on InP substrates. (a) As-grown InAs nanowire arrays on InP (111)B substrate. (b) Schematics of the unit cell of an InAs nanowire photodetector. (InP passivation layer is not shown.) (c) Close-up view of plasmonic gratings. (d) Wire-bonded photodetector device sample (left). (d) Close-up view of the wire-bonded nanowire array (right). The size of the array is 100 μm × 100 μm. (e) Spectral response of the InAs photodetector at a reverse bias of 0.5 V, indicating photodetection signature at SWIR up to 2.5 μm. The inset shows the calculated EQE in fractions. Ref. 13.

In this work, we have developed a comprehensive 3D computational model, which accurately simulates photocurrent and external quantum efficiency (EQE) of these "small" photodetectors [7-9]. The model was successfully implemented to design nanowire-based InAs(Sb) infrared photodetectors, where each nanowire array composed of thousands of nanowires is a single pixel. InAs(Sb) nanowires are grown by selective-area metal-organic chemical vapor deposition [10-12]. The This further leads to the development of uncooled nanowire photodetection platforms from short-wave infrared (SWIR) (Figure 1) to mid-wave infrared (MWIR) (Figure 2), spanning 1.4µm – 5µm [13,14]. Note that the design of pixels in sub-10 micron or sub-micron scale is extremely difficult, much more than that of bulk detectors in hundreds of micron. Their 3D features need to fully accounted for in simulation and cannot be simplified unlike planar mesa structures. Therefore, analytical solutions are of limited applicability, and it is necessary to use a 3D computational model.

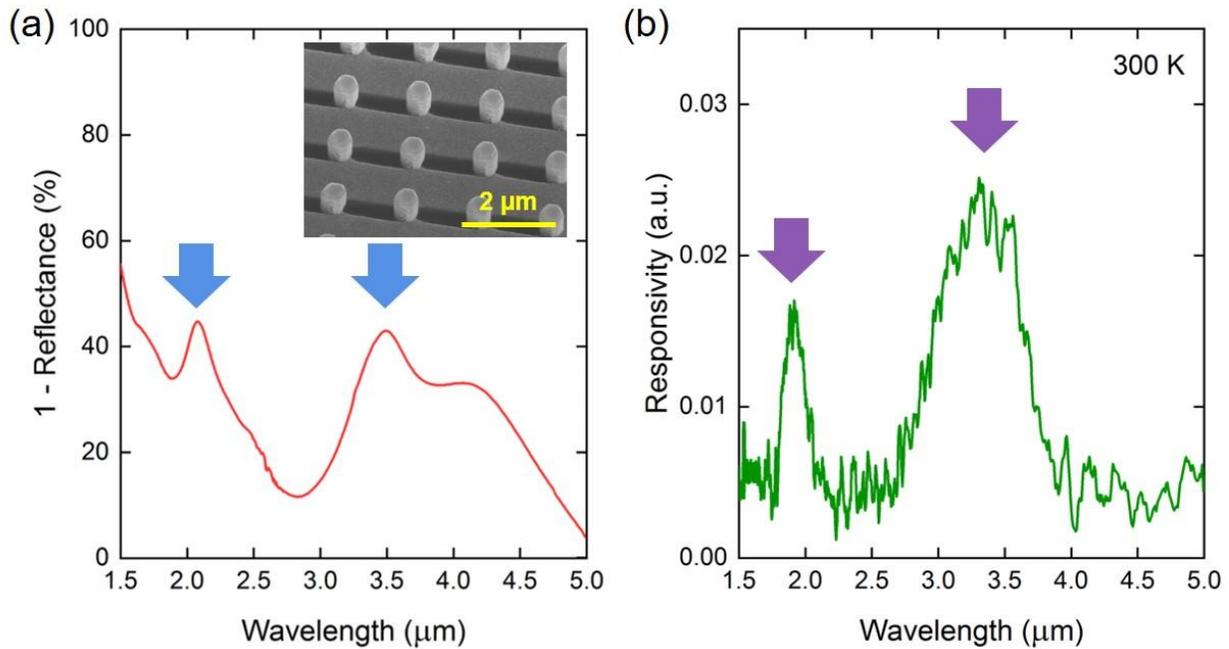

**Figure 2.** MWIR InAsSb nanowire detectors on InP substrates. (a) Reflectance measurement. The yaxis is one minus reflectance, giving the sum of absorptance and transmittance. The inset shows the top view of the fabricated nanowire−plasmonic array. (b) Spectral response measurement at room temperature (not normalized). The arrows highlight the peaks observed in both reflectance and spectral response measurement. Ref. 14.

This computational model combines optical simulation by time-domain finite-difference (FDTD) and electrical simulation by finite-element method (FEM) (Figure 3) [7,8]. First, the 3D electric field profile and optical absorption are solved by FDTD. Then, the simulated electric field and optical generation profiles in FDTD mesh grids are interpolated into the FEM grids so that it could be imported into the electrical simulator. Finally, the imported electric field profiles were converted to optical generation profiles and the electrical simulator solved the drift-diffusion and continuity equations to output the photo current. As a result, we can couple simulation results between numerical techniques and solver platforms that are individually optimized for optical and electrical simulations. This method allows to determine and optimize our nanowire device performance metrics. It is the first computational model which can simulate photocurrent from 3D optical profiles. The early version of the model was used to simulate temporal carrier motions in 3D nanowires. It was later optimized the model by including more complicated carrier dynamics and material properties.

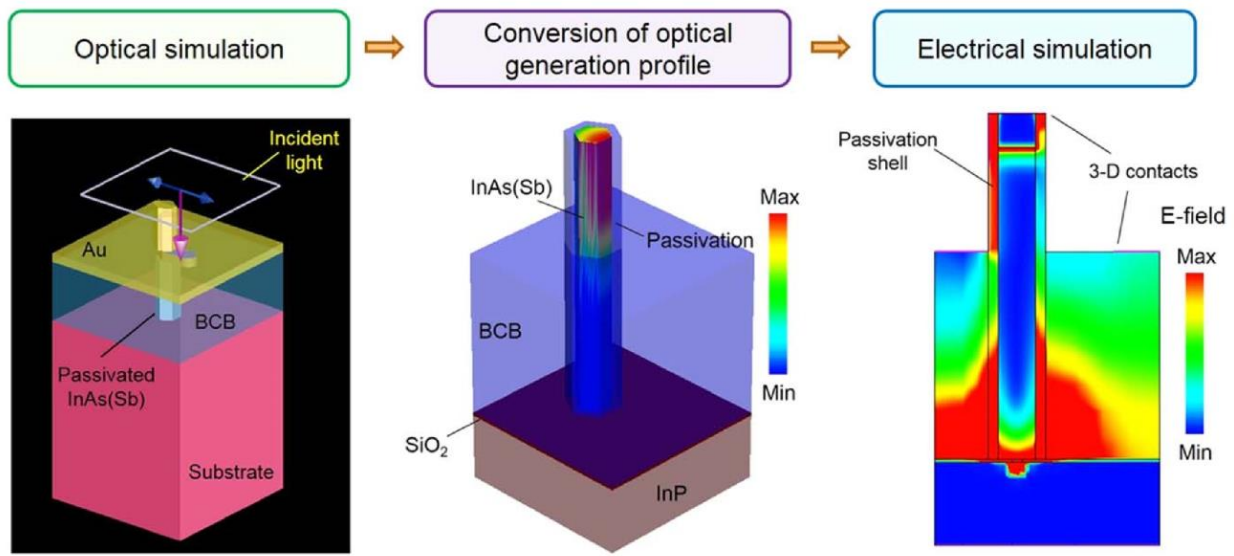

**Figure 3.** Schematic diagram of simulation process, which is composed of three major steps: (1) optical simulation by FDTD, (2) conversion of optical profile from FDTD mesh to FEM mesh, and (3) electrical simulation by FEM. E-field in (3) represents the device E-field profile. Ref. 7.

Guided by the 3D computational model (Figure 4), we demonstrated two prototypes – uncooled InAs nanowire detector at SWIR with a cut-off at 2.5 µm and InAsSb nanowire detector at MWIR up to 3.4µm [13,14]. The nanowire photodetector arrays were grown by selective-area MOCVD with extremely high vertical yield and high uniformity. We further studied the impact of individual material properties, e.g. surface recombination and carrier mobilities, on photoresponse. This capability allows to concurrently explore multiple material properties that reflect the complex carrier dynamics in 3D detector structures.

Furthermore, the unique plasmonic structure in our devices also opens the possibility of developing nanowire-plasmonic multichannel detectors for multispectral and hyperspectral imaging applications.

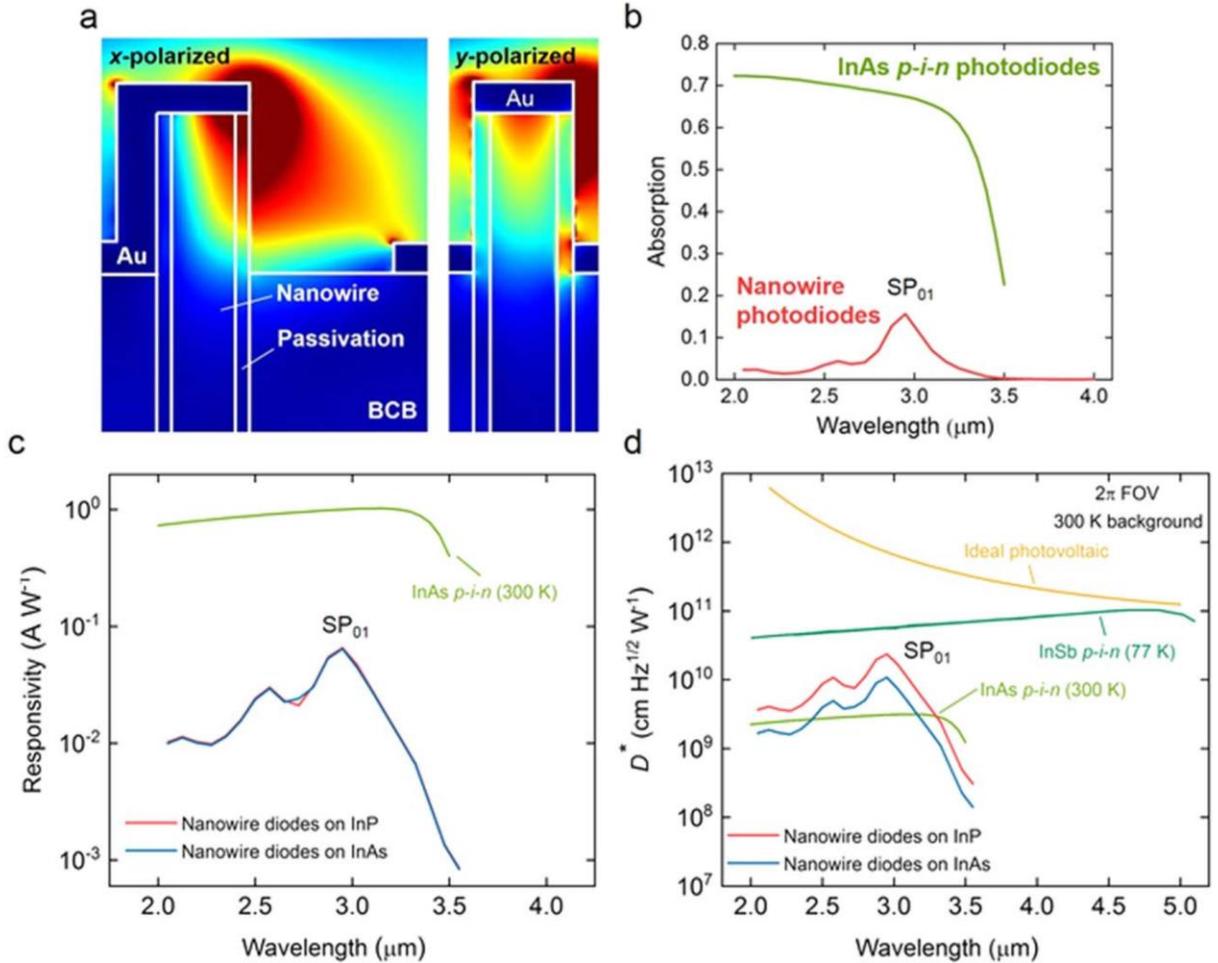

**Figure 4.** Simulated spectral response of InAs(Sb) nanowire photodiodes. (a) Optical generation: *x*-polarized and *y*-polarized electric profiles at 3 μm. (b) Optical absorption of InAsSb nanowire photodiodes and planar InAs photodiodes. (c) Responsivity of nanowire photodiodes on InP and InAs substrates. (d) $D^*$ of nanowire photodiodes on InP and InAs substrates, where $D^*$ from reference cells of commercial uncooled InAs photodiodes and cryogenically cooled (77 K) InSb photodiodes are also presented. Ref. 7.

In conclusion, we show that the 3D computational model can support sophisticated optical and electrical design of nanostrucutred photodetectors to achieve better performance. We believe this study paves the way for designing next-generation semiconductor-based nanostructured optoelectronic devices.